\documentclass[
amsmath,
amssymb,
aps, 
pra,
twocolumn, 
groupedaddress,
nobalancelastpage,
floatfix]
{revtex4-2}
\usepackage{mathtools}
\usepackage[colorlinks]{hyperref}
\usepackage{lipsum}
\usepackage{bbold}
\usepackage{bm}
\usepackage{dsfont}
\usepackage{enumerate}
\usepackage{graphicx}
\usepackage{float}
\usepackage {xcolor}
\usepackage{cleveref}
\usepackage{amsmath}

\DeclareMathOperator{\e}{e}					

\newcommand{\nullarg}{{}\cdot{}}

\renewcommand{\vec}[1]{\bm{#1}}	

\DeclarePairedDelimiterX{\mean}[1]{\langle}{\rangle}{
	\ifblank{#1}{\nullarg}{#1}
}
\DeclarePairedDelimiterX{\abs}[1]{\lvert}{\rvert}{
	\ifblank{#1}{\nullarg}{#1}
}
\DeclarePairedDelimiterX{\norm}[1]{\lVert}{\rVert}{
	\ifblank{#1}{\nullarg}{#1}
}
\DeclarePairedDelimiterX{\bra}[1]{\langle}{\rvert}{#1}

\DeclarePairedDelimiterX{\ket}[1]{\lvert}{\rangle}{#1}
\DeclarePairedDelimiterX{\braket}[2]{\langle}{\rangle}{
	\ifblank{#1}{\nullarg}{#1} \delimsize\vert \ifblank{#2}{\nullarg}{#2}
}
\DeclarePairedDelimiterX{\sandwich}[3]{\langle}{\rangle}{
	\ifblank{#1}{\nullarg}{#1} \delimsize\vert \ifblank{#2}{\nullarg}{#2} \delimsize\vert \ifblank{#3}{\nullarg}{#3}
}
\DeclarePairedDelimiterX{\inner}[2]{\langle}{\rangle}{
	\ifblank{#1}{\nullarg}{#1} , \ifblank{#2}{\nullarg}{#2}
}

\begin{document}
	
	\title{Hierarchy of approximations for describing quantum light from high-harmonic generation: A Fermi-Hubbard model study}
	
	\author{Christian Saugbjerg Lange}
	\author{Lars Bojer Madsen} 
	\affiliation{Department of Physics and Astronomy, Aarhus University, Ny Munkegade 120, DK-8000 Aarhus C, Denmark}
	\date{\today}
	\begin{abstract}
		The quantum optical description of high-order harmonic generation where both the electrons of the generating medium and the driving and generated light fields are described quantum mechanically has been of significant interest in the past years. The quantum optical formulation leads to equations of motion for the generated light field in which the quantum optical field couples to the time-dependent current of the electronic medium irrespectively of the specifics of the electronic system being an atom, molecule, or solid. These equations of motion are not solvable for any realistic system and accurate and verified approximations are hence needed. In this work, we present a hierarchy of approximations for the equations of motion for the photonic state. At each level in this hierarchy, we compare it to the previous level justifying the validity using the Fermi-Hubbard model as an example of an electronic system with correlations. This model allows us to perform an accurate simulation of the electron motion of all the required states. We find that for the typical experimental situation of  weak quantized-light-matter-coupling constant and at intensities well below the damage threshold, an explicit expression for the generated quantum light, referred to as the Markov-state approximation (MSA), captures the high-harmonic spectrum quantitatively and describes the single-mode quantum properties of the generated light as characterized by the Mandel-Q parameter and the degree of squeezing qualitatively.
	\end{abstract}
	
	\maketitle
	\newpage
	\section{Introduction}
	High-harmonic generation (HHG) is a highly nonlinear process where an intense laser field interacts with a quantum system (atoms, molecules, solids) and an upconversion of the laser frequency occurs, resulting in the emission of light consisting of higher harmonics of the driving field. For decades, a semiclassical description of HHG has proven successful in its predictability and description of strong-field phenomena and attosecond physics \cite{Krausz2009}. This semiclassical description of HHG, where the electronic system is described quantum mechanically and the involved light fields are described classically, has successfully predicted observables such as the harmonic cutoff, selection rules for the presence of even and odd harmonics, and their polarization \cite{Neufeld2019}.

	Though the semiclassical description of HHG has proven useful for a wide range of applications, it cannot account for the quantum optical nature of neither the driving field nor the emitted light. In recent years, there has been a growing interest in describing HHG from a fully quantum perspective, i.e., with both a quantized electronic system and a quantum optical description of the involved light fields. Theoretical work using a coherent driving field has studied both atomic gasses \cite{Gorlach2020_NatComm, Yangaliev2020_OlegToltikhin, LewensteinCatState2021, Gombkoto2021, RiveraDean2022, Yi2024, Stammer2024_MarkovPaper}, molecules \cite{RiveraDean2024_H2_molecule}, semiconductors \cite{Riveradean2023entanglementHHG, RiveraDean2024_Bloch_based, Gonoskov2024}, and correlated systems \cite{Pizzi2023_KaminerManyBody, Lange2024_electroncorrelation}. Further, using intense nonclassical states of light as driving fields has been considered \cite{Tzur2023_photon_statistics_force, Gorlach2023_HHG_driven_by, Tzur2024_generation_of_squeezed, Stammer2024_absence_of_qo_coherence}, highlighting the various ways of engineering nonclassical states of light from high-harmonic generation.
	
	A seminal experimental report of nonclassical light from HHG was on the so-called 'Cat-state', a macroscopic superposition between two different coherent states, which was created with a post-selection measurement scheme of the emitted light \cite{LewensteinCatState2021}. Recent experimental work has reported that HHG can indeed be nonclassical \cite{Theidel2024_Biegert_Experiment} without any post selection and also considered nonclassical driving fields \cite{Rasputnyi2024, Lemieux2024} showing how the nonclassicality of the driving field is transferred to some of the emitted harmonics. 
	
	Central to the understanding of all of the above approaches is the quantum optical description of an electronic system driven by a coherent-state laser. Interestingly, there are many open questions in this direction. One such open question is the validity of typical approximations used. As the exact equations of motion for the quantum optical state of the emitted light are not solvable in the general case, valid approximations are called for, potentially improving the analytical understanding. In this work, we introduce a hierarchy of approximations to the equations of motion for the emitted photonic state similar to those considered in Ref. \cite{Stammer2024_MarkovPaper} where each step builds on top of the previous approximations. Here, each step is numerically verified using the Fermi-Hubbard model, a generic many-electron model including electron correlations, which allows for an accurate solution of the electronic problem as all electronic states can be included for a suitable size chain. This model, hence, allows for a quantitative assessment of all levels of approximations and qualifies a systematic discussion of these. Working within these justified approximations, we show how the nonclassical features (photon statistics, photon squeezing) of the emitted light relate to the transition current elements of the classically driven system or, equivalently, to the time correlations of the classical current. As these transition currents do not, in general, have the same spectral features as the HHG spectrum \cite{Lange2024_electroncorrelation}, this explains why the nonclassical features peak in signal at other frequencies than the peaks in the spectrum as found in Refs. \cite{Gorlach2020_NatComm, Lange2024_electroncorrelation}. Employing these validated approximations simplifies and reduces the equations to be solved, possibly allowing simulation of larger systems in the future. 
	In particular, we show that an explicit closed expression for the emitted quantum light, obtained within what we call the Markov-state approximation (MSA) (see also Ref. \cite{Stammer2024_MarkovPaper}), captures qualitatively both the shape of HHG spectra and quantum observables such as the Mandel-Q parameter and the squeezing of the emitted light.

	This paper is organized as follows. First, an introduction to quantum optical HHG is given in Sec. \ref{sec:theory}. Then, a hierarchy of approximations is introduced in Sec. \ref{sec:hierarchy_of_approx} followed by a specification of the electronic system in Sec. \ref{sec:electronic_system}. In Sec. \ref{sec:results_and_discussion}, a presentation of numerical results is given followed by a discussion of the validity of the approximations. Finally, a conclusion and an outlook is given in Sec. \ref{sec:conclusion_and_outlook}.

	\section{Theory} \label{sec:theory}
	\subsection{Quantum optical description of high-harmonic generation}
	In this section, we briefly present the key equations of high-harmonic generation from a quantum optical perspective. For related detailed derivations see, e.g., Refs. \cite{Lange2024_electroncorrelation, Gorlach2020_NatComm, Stammer2023_Lewenstein_tutorial_Quantum_state_engineering}. We consider the time-dependent Schrödinger equation (TDSE) [atomic units are used throughout]
	
	\begin{equation}
		i \frac{\partial}{\partial t} \ket{\Psi(t)} = \hat{H} \ket{\Psi(t)}, \label{eq:TDSE}
	\end{equation}
	where $\ket{\Psi(t)}$ is the combined state of both the electronic and photonic system and where
	\begin{equation}
		\hat{H} = \dfrac{1}{2} \sum_{j=1}^N (\hat{\vec{p}}_j + \hat{\vec{A}})^2 + \hat{U} + \hat{H}_F \label{eq:Hamiltonian_most_general}
	\end{equation}
	is the general many-electron Hamiltonian consisting of $N$ electrons. Here, $\hat{\vec{p}}_j$ is the momentum for the electron with the index $j$, $\hat{U}$ is accounting for the electron-electron and electron-nuclear interaction,

	\begin{equation}
		\hat{\vec{A}} = \sum_{\vec{k}, \sigma} \dfrac{g_0}{\sqrt{\omega_k}} (\hat{\vec{e}}_\sigma \hat{a}_{\vec{k}, \sigma}+ \hat{\vec{e}}_\sigma^* \hat{a}_{\vec{k}, \sigma}^\dagger) \label{eq:Vector_potential_quantized}
	\end{equation}
	is the quantized vector potential in the dipole approximation, and $\hat{H}_F = \Sigma_{\vec{k}, \sigma} \omega_k \hat{a}^\dagger_{\vec{k}, \sigma} \hat{a}_{\vec{k}, \sigma}$ is the Hamiltonian of the free electromagnetic field. In Eqs. (\ref{eq:Hamiltonian_most_general}) and (\ref{eq:Vector_potential_quantized}) the sum $\Sigma_{\vec{k}, \sigma}$ is over all photonic wavenumbers $\vec{k}$ and polarizations $\sigma$ with unit vector $\hat{\vec{e}}_\sigma$, and 	$g_0 = \sqrt{2\pi/V}$ is the effective coupling for the quantization volume $V$. The operator $\hat{a}_{\vec{k}, \sigma}$ ($\hat{a}_{\vec{k}, \sigma}^\dagger$) annihilates (creates) a photon with the frequency $\omega_k = \lvert \vec{k} \rvert c$.
	The initial state of the system prior to any interaction between the laser and electrons is $\ket{\Psi(t)} = \ket{\phi_i} \ket{\psi_{laser}(t)}$, where $\ket{\phi_i}$ is the initial field-free electronic eigenstate (which we take to be the ground state) and 
	\begin{align}
		\ket{\psi_{laser}(t)} =  \bigotimes_{\vec{k}_L, \sigma_L} \hat{\mathcal{D}}(\alpha_{\vec{k}_L, \sigma_L}(t)) \ket{0}
	\end{align}
	is a multimode coherent state involving the laser modes ($\vec{k}_L, \sigma_L$) and where
	\begin{equation}
		\hat{\mathcal{D}}(\alpha_{\vec{k}, \sigma}(t)) = e^{\alpha_{\vec{k}, \sigma}(t) \hat{a}_{\vec{k}, \sigma}^\dagger  -\alpha_{\vec{k}, \sigma}^*(t) \hat{a}_{\vec{k}, \sigma}}
	\end{equation}
	is the displacement operator with coherent state amplitude $\alpha_{\vec{k}, \sigma}(t) = \alpha_{\vec{k}, \sigma} \exp{(-i\omega_k t)}$.
			
	To avoid dealing with a macroscopic number of photons, we transform away the driving field of the laser and the Hamiltonian in Eq. (\ref{eq:Hamiltonian_most_general}) separates into $\hat{\tilde{H}}(t) = \hat{H}_{sc}(t) + \hat{V} + \hat{H}_F$, where the tilde denotes a displaced frame, $\hat{V}$ is the coupling between the electrons and the quantized field, and $\hat{H}_{sc}(t)$ is the semiclassical Hamiltonian similar to Eq. (\ref{eq:Hamiltonian_most_general}) but where $\hat{\vec{A}} \rightarrow \vec{A}_{cl}(t)$, i.e., the quantized vector potential has been replaced by a classical driving potential. Going into a rotating frame with respect to both $\hat{H}_{sc}(t)$ and $\hat{H}_F$, we obtain the equation

	\begin{equation}
		i \dfrac{\partial}{\partial t} \ket{\tilde{\Psi}(t)}_I = \hat{V}_I(t) \ket{\tilde \Psi(t)}_I,
	\end{equation} 
	where the subscript $I$ indicates that the state is in the rotating frame. The interaction between the quantized vector potential and the electrons is given by the coupling 
	 \begin{equation}
	 	\hat{V}_I(t) = \hat{\vec{A}}_{Q}(t) \cdot \sum_{m,n} \vec{j}_{m,n}(t) \ket{\phi_m} \bra{\phi_n}, \label{eq:general_interaction}
	 \end{equation}
	 with $\ket{\phi_m}$ and $\ket{\phi_n}$ denoting time-independent eigenstates of the field-free Hamiltonian for the electronic system, and with the time-dependent quantized vector potential
	 \begin{equation}
	 	\hat{\vec{A}}_{Q}(t) = \sum_{\sigma, \vec{k}} \dfrac{g_0}{\sqrt{\omega_k}} (\hat{\vec{e}}_\sigma \hat{a}_{\vec{k}, \sigma}  \e^{-i \omega_k t}+ \hat{\vec{e}}_\sigma^* \hat{a}_{\vec{k}, \sigma}^\dagger \e^{i \omega_k t}), \label{eq:Vector_potential_dipole_time_dep}
	 \end{equation}
	 and the transition current matrix elements
	 \begin{equation}
	 	\vec{j}_{m,n}(t) = \bra{\phi_m(t)} \hat{j}(t) \ket{\phi_n(t)}, \label{eq:current_elements}
	 \end{equation}
	 where $\hat{\vec{j}}(t) = \Sigma_{j = 1}^N [\hat{\vec{p}}_j + \vec{A}_{cl}(t)]$ is the electronic current operator. In Eq. (\ref{eq:current_elements}),  $\ket{\phi_m(t)}$ is the time-evolved $m$'th field-free eigenstate satisfying the TDSE with the semiclassical Hamiltonian
	 \begin{equation}
	 	i \dfrac{\partial}{\partial t} \ket{\phi_m(t)} = \hat{H}_{sc}(t) \ket{\phi_m(t)}. \label{eq:TDSE_SC}
	 \end{equation}
	 
	 By expanding the full state of the combined electronic and photonic system in terms of field-free electronic eigenstates
	 \begin{equation}
	 	\ket{\tilde{\Psi}(t)}_I = \sum_m \ket{\phi_m} \ket{\chi^{(m)}(t)},
	 \end{equation}
	 and projecting onto the electronic state $\bra{\phi_m}$ we obtain the equation of motion for the corresponding photonic state 
	 \begin{equation}
	 	i \dfrac{\partial}{\partial t} \ket{\chi^{(m)}(t)} = \hat{\vec{A}}_Q(t) \cdot \sum_n \vec{j}_{m,n}(t) \ket{\chi^{(n)}(t)}, \label{eq:chi_mn_full}
	 \end{equation}
	Equation (\ref{eq:chi_mn_full}) is the central equation describing the evolution of the quantized photonic state and is describing the coupling between the quantized vector potential and the transition current elements between different dressed electronic states. As such, we emphasize that Eq. (\ref{eq:chi_mn_full}) is completely general and does not depend on the electronic medium (atom, molecule, solid). Equation (\ref{eq:chi_mn_full}) is also found in, e.g., Refs. \cite{Gorlach2020_NatComm, Lange2024_electroncorrelation} and in a length gauge formulation in Refs. \cite{Yi2024, RiveraDean2022, RiveraDean2024_H2_molecule, Stammer2024_MarkovPaper}. Unfortunately, the size of the combined Hilbert space of both photons and electrons generally impedes any direct numerical solution of Eq. (\ref{eq:chi_mn_full}) without further approximations. These will be introduced and discussed in Sec. \ref{sec:hierarchy_of_approx}. First, we turn to a presentation of the observables of interest in the present study.
	
	\subsection{Observables}
	In the quantum optical formalism, the HHG spectrum is given by \cite{Lange2024_electroncorrelation}
	\begin{equation}
		S(\omega_k) = \dfrac{\omega_k^3}{g_0^2 (2\pi)^2 c^3} \sum_{\sigma} \langle \hat{n}_{\vec{k}, \sigma} \rangle,  \label{eq:spectrum_qo}
	\end{equation}
	where $\hat{n}_{\vec{k}, \sigma} = \hat{a}_{\vec{k}, \sigma}^\dagger \hat{a}_{\vec{k}, \sigma}$ is the photonic counting operator. This is different from the usual semiclassical spectrum given by \cite{Baggesen2011}
	
	\begin{equation}
		S_{cl}(\omega) = \omega^2 \lvert \tilde{\vec{j}}_{i,i}(\omega) \rvert^2, \label{eq:spectrum_sc}
	\end{equation}
	where $ \tilde{\vec{j}}_{i,i}(\omega)$ is the Fourier transform of the classical current.
	
	In the quantum optical description of HHG, however, more than just the spectrum can be measured. In particular, the photon statistics and squeezing properties of the generated light are of great interest with regard to nonclassical features.
	The photon statistics can be quantified by the Mandel-Q parameter defined by \cite{Gerry_knight_2004}
	\begin{equation}
		Q_{\vec{k}, \sigma} = \dfrac{\langle  \hat{n}_{\vec{k}, \sigma}^2 \rangle - \langle  \hat{n}_{\vec{k}, \sigma}  \rangle^2 }{\langle  \hat{n}_{\vec{k}, \sigma}  \rangle}  -1, \label{eq:mandel_Q_def}
	\end{equation}
	for a given mode. A classical coherent state will have Poissonian statistics and hence $Q_{\vec{k}, \sigma} = 0$.  If $Q_{\vec{k}, \sigma}>0$ the photon emission follows super-Poissonian statistics while $Q_{\vec{k}, \sigma}<0$ yields sub-Poissonian statistics, corresponding to a photon number distribution wider or narrower than a Poissonian distribution, respectively. While both a classical and nonclassical state can yield super-Poissonian statistics, sub-Poissonian statistics is a clear telltale sign of a nonclassical state \cite{Gerry_knight_2004}. Another telltale sign of nonclassicality is a nonvanishing squeezing  \cite{scully_zubairy_1997, Braunstein2005}. The degree of squeezing can be quantified by the squeezing parameter $\eta_{\vec{k}, \sigma}$. In the unit of dB, $\eta_{\vec{k}, \sigma}$ is given as
	\begin{equation}
		\eta_{\vec{k}, \sigma} = - 10 \log_{10} \big\{4 \underset{\theta \in [0, \pi)}{\text{min}} [\Delta \hat{X}_{\vec{k}, \sigma}(\theta)]^2\big\}, \label{eq:squeezing_def}
	\end{equation}
	where $\hat{X}_{\vec{k}, \sigma}(\theta) = (\hat{a}_{\vec{k}, \sigma} e^{-i \theta} + \hat{a}_{\vec{k}, \sigma}^\dagger e^{i \theta})/2$ is the quadrature operator. The angle, $\theta$, that minimizes the variance of the quadrature operator in Eq. (\ref{eq:squeezing_def}) gives the direction in phase space where the uncertainty in the corresponding quadrature is decreased below that of a coherent state at the expense of increasing the uncertainty in the conjugate quadrature. For a coherent state $\eta_{\vec{k}, \sigma} = 0$ for all polarizations and modes. In Sec. \ref{sec:results_and_discussion}, we consider results for the spectra [Eq. (\ref{eq:spectrum_qo})], the Mandel-Q parameter [Eq. (\ref{eq:mandel_Q_def})], and the squeezing parameter [Eq. (\ref{eq:squeezing_def})].

	\section{Hierarchy of Approximations} \label{sec:hierarchy_of_approx}
	The exact equation of motion for the photonic state [Eq. (\ref{eq:chi_mn_full})] is not solvable for any realistic system, as the number of photonic states required would be too large when expanded in, e.g., a Fock basis
	\begin{equation}
		\ket{\chi^{(m)}(t)} = \sum_{\{n \}} c_{\{n\}}^{(m)}(t) \ket{n_{\vec{k}_1, \sigma_,1}, n_{\vec{k}_2, \sigma_2}, \dots } \label{eq:chi_m_exact_expansion},      
	\end{equation} 
	where the sum is over all $\{n\}$ possible combinations of photon numbers in all the considered modes. The number of basis states required in Eq. (\ref{eq:chi_m_exact_expansion}) is $p^{\vec{k}_{max}}$, where $p$ is the maximum number of photons allowed in a given mode, and $\vec{k}_{max}$ is the highest mode considered. For an extended HHG spectrum, this basis is too large even when allowing for only a few photons per mode, and hence further approximations on the photonic state are necessary. Thus far, two different approaches have been taken in the literature, as we now summarize.
	
	One approach is to neglect all transition currents generated by $\ket{\phi_m(t)}$ for $m\neq i$ in Eq. (\ref{eq:chi_mn_full}) and only keep $\vec{j}_{i,i}(t) = \bra{\phi_i(t)} \hat{\vec{j}}(t) \ket{\phi_i(t)}$, i.e., the current generated by the solution to the semiclassical TDSE, when the electronic system starts in the ground state $\ket{\phi(t=0)} = \ket{\phi_i}$. This limit yields an analytical solution for the photonic state as a product of coherent states \cite{LewensteinCatState2021, RiveraDean2022, Stammer2023_Lewenstein_tutorial_Quantum_state_engineering, RiveraDean2024_Bloch_based}, i.e., the most classical state possible. From these states, one can perform a conditioning measurement to generate cat states \cite{LewensteinCatState2021, Stammer2023_Lewenstein_tutorial_Quantum_state_engineering, RiveraDean2022}. Other works have included both the ground state and an excited state (possibly resonantly coupled by the laser) in Eq. (\ref{eq:chi_mn_full}), and under certain approximations an expression for the photonic quantum state associated with the two bound states can be obtained \cite{RiveraDean2024_H2_molecule, Yi2024, RiveraDean2024_Squeezed_states_of_light}. This approach neglects the contribution from many bound and continuum states which will affect the quantum properties of the emitted light as detailed below. In contrast to the case where only the ground state is considered, this inclusion of more than a single photonic state, however, allows for nondiagonal transition currents in Eq. (\ref{eq:chi_mn_full}) which is, as we stress below, a key ingredient towards generating nonclassical light in the HHG process itself. We further note that the required transition current elements in Eq. (\ref{eq:chi_mn_full}), e.g. $\vec{j}_{i,i}(t)$, typically are obtained with the inclusion of more electronic states, either from an exact integration of the TDSE or via the strong-field approximation (SFA) ansatz for the wave function. 
		
\begin{figure}
		\centering
		\includegraphics[width=1.0\linewidth]{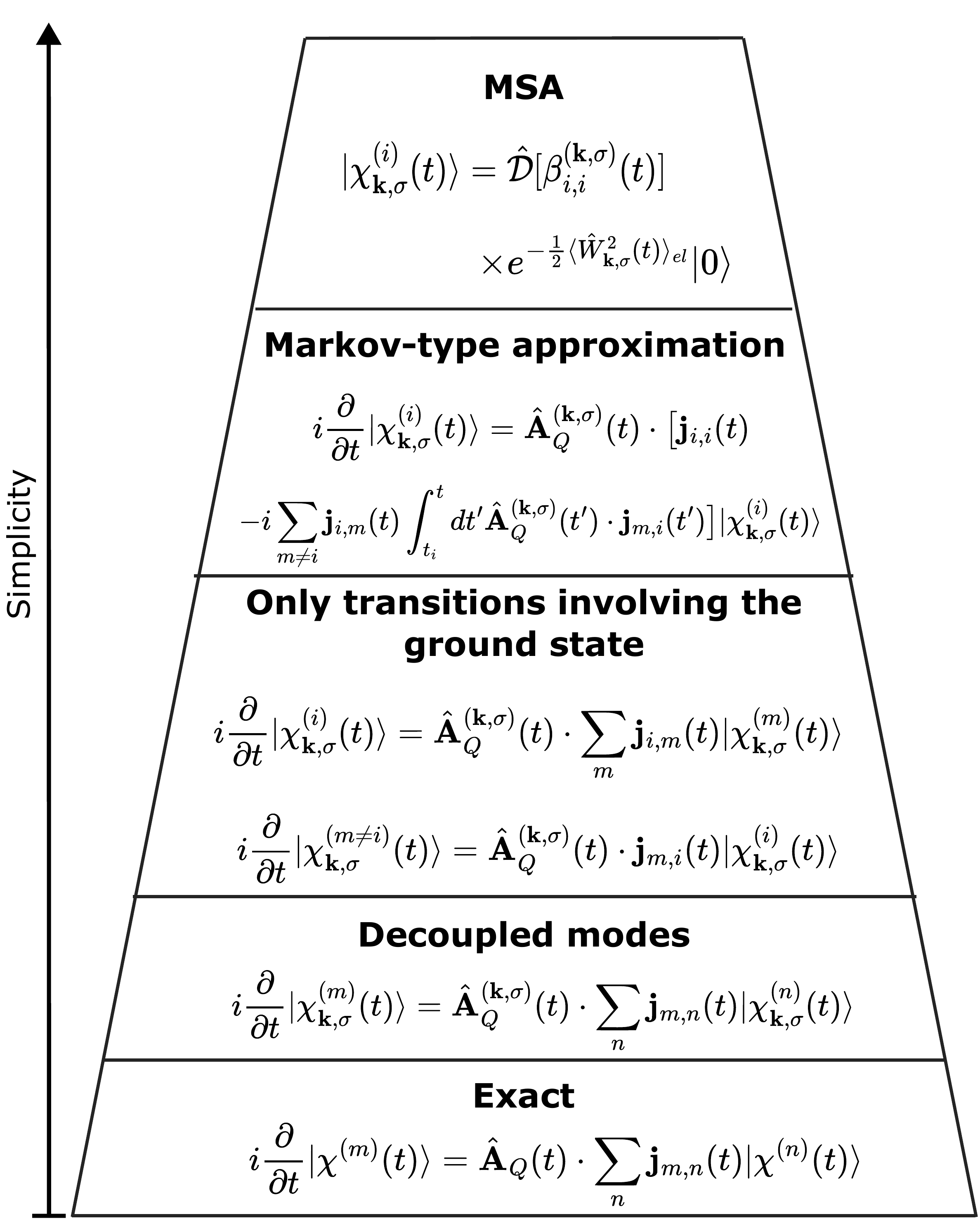}
		\caption{Hierarchy of approximations with related equations for the photonic states as discussed in detail in the text. With increasing simplicity, each approximation assumes all the previous approximations made.}
		\label{fig:hierachyofapproximations}
	\end{figure}

	Another approach to simplify Eq. (\ref{eq:chi_mn_full}) is to decouple the photonic modes such that each mode is solved independently \cite{Gorlach2020_NatComm, Lange2024_electroncorrelation}. Different to the first essential state approach, this allows one to keep all photonic states in Eq. (\ref{eq:chi_mn_full}) thus keeping contributions from all transition currents. In terms of the ansatz for the state in mode ($\vec{k}, \sigma$), Eq. (\ref{eq:chi_m_exact_expansion}) reduces, in the decoupled case, to
	\begin{equation}
		\ket{\chi^{(m)}_{\vec{k}, \sigma}(t)} = \sum_n c_n^{(m)}(t) \ket{n_{\vec{k}, \sigma}},
	\end{equation}
	which requires only $p \times \vec{k}_{max}$ basis states for a given state with index $m$, drastically reducing the computational requirements. On the other hand, this approximation to decouple the photonic modes neglects some quantum features such as two-mode squeezing and entanglement between different harmonics. For the rest of this work, we follow this latter decoupled approach as the two-mode photonic coupling is insignificant for most pairs of harmonic modes as indicated in Ref. \cite{RiveraDean2024_Squeezed_states_of_light}.
	In the remainder of this section we will derive a hierarchy of approximations where each new approximation assumes the previous approximations. These approximations are made to both ease the numerical cost required to simulate the system while also, at the same time, providing a better analytical understanding of the underlying physics. The hierarchy of the approximations made in this section and their related equations are summarized in Fig. \ref{fig:hierachyofapproximations} and will be evaluated and discussed further in Sec. \ref{sec:results_and_discussion}.
	
	\subsection{Decoupled modes}
		To proceed from Eq. (\ref{eq:chi_mn_full}), we neglect all couplings between different modes and solve the decoupled system of equations as done in Refs. \cite{Gorlach2020_NatComm, Lange2024_electroncorrelation}.  The state $ \ket{\chi^{(m)}(t)}$ in Eq. (\ref{eq:chi_mn_full}) is then approximated as a product state
		\begin{equation}
			 \ket{\chi^{(m)}(t)} = \otimes_{\vec{k}, \sigma}  \ket{\chi_{\vec{k}, \sigma}^{(m)}(t)}, \label{eq:chi_mn_product_state}
		\end{equation}
		and each state on the right-hand side of Eq. (\ref{eq:chi_mn_product_state}) evolves according to 
		\begin{equation}
			i \dfrac{\partial}{\partial t} \ket{\chi_{\vec{k}, \sigma}^{(m)}(t)} = \hat{\vec{A}}_Q^{(\vec{k}, \sigma)}(t) \cdot \sum_n \vec{j}_{m,n}(t) \ket{\chi_{\vec{k}, \sigma}^{(n)}(t)}, \label{eq:chi_mn_decoupled}
		\end{equation}
		where the quantized field operator $\hat{\vec{A}}_Q^{(\vec{k}, \sigma)}(t)$ only acts on mode ($\vec{k}, \sigma$)
		\begin{equation}
			\hat{\vec{A}}_Q^{(\vec{k}, \sigma)}(t) = \dfrac{g_0}{\sqrt{\omega_k}} ( \hat{\vec{e}}_\sigma \hat{a}_{\vec{k}, \sigma} e^{-i \omega_k t}+  \hat{\vec{e}}_\sigma^* \hat{a}_{\vec{k}, \sigma} ^\dagger e^{i \omega_k t}). \label{eq:AQ_single_mode}
		\end{equation}
		Note that solving Eq. (\ref{eq:chi_mn_decoupled}) will yield the most exact practically obtainable results within this framework. 
	
	\subsection{Keeping only transitions involving the electronic ground state}
	As a next approximation, we include only couplings that involve the electronic ground state, i.e., we only include transition current elements $\vec{j}_{i,m}(t)$ and $\vec{j}_{m,i}(t)$ [see Eq. (\ref{eq:current_elements})], where the subscript $i$ denotes the ground state, and neglect all other terms in Eq. (\ref{eq:chi_mn_decoupled}). As the initial photonic state is the vacuum state associated with the field-free electronic ground state, $c_n^{(m)}(0) = \delta_{m,i} ~\delta_{n,0}$ [see Eq. (\ref{eq:chi_mn_product_state})], the coupling to a state $ \ket{\chi^{(m)}(t)}$ is via $	\hat{\vec{A}}_Q^{(\vec{k}, \sigma)} \cdot \vec{j}_{i,m}(t)$, as seen in Eq. (\ref{eq:chi_mn_decoupled}), which is first order in the coupling constant $g_0$. As we take $g_0 = 4 \times 10^{-8}$ a.u. (similar to Refs. \cite{LewensteinCatState2021, Lange2024_electroncorrelation, Tzur2024_generation_of_squeezed}, see also \cite{Riek2015}), this coupling is weak and we neglect all higher-order couplings which are those that do not involve the initial state. From Eq. (\ref{eq:chi_mn_decoupled}), we hence obtain the approximate set of equations
	\begin{subequations} \label{eq:chi_im}
			\begin{align}
			i \dfrac{\partial}{\partial t} \ket{\chi_{\vec{k}, \sigma}^{(i)}(t)} & = \hat{\vec{A}}_Q^{(\vec{k}, \sigma)}(t) \cdot \sum_{m} \vec{j}_{i,m}(t) \ket{\chi_{\vec{k}, \sigma}^{(m)}(t)}, \label{eq:chi_im_initial}  \\
			i \dfrac{\partial}{\partial t} \ket{\chi_{\vec{k}, \sigma}^{(m \neq i)}(t)} & = \hat{\vec{A}}_Q^{(\vec{k}, \sigma)}(t) \cdot  \vec{j}_{m,i}(t) \ket{\chi_{\vec{k}, \sigma}^{(i)}(t)} \label{eq:chi_im_other}.
		\end{align}
	\end{subequations}
	We note that Eq. (\ref{eq:chi_im}) only requires $M$ transition current elements $\{\vec{j}_{i,m}(t)\}$, while the full system of equations in Eq. (\ref{eq:chi_mn_decoupled}) requires $M^2$ transition current elements $\{\vec{j}_{m,n}(t)\}$, with $M$ being the total number of field-free electronic eigenstates in the simulation. As such the memory requirement is less demanding and larger systems might be considered. 
	\subsection{Markov-type approximation}
	To go further, we seek to bring Eq. (\ref{eq:chi_im}) into a simpler form. We first formally integrate Eq. (\ref{eq:chi_im_other}) 
	\begin{equation}
		\ket{\chi^{(m\neq i)}_{\vec{k}, \sigma}(t)} = - i \int_{t_0}^{t} dt' \hat{\vec{A}}_Q^{(\vec{k}, \sigma)}(t') \cdot  \vec{j}_{m,i}(t') \ket{\chi_{\vec{k}, \sigma}^{(i)}(t')}. \label{eq:chi_mi_formal_integration}
	\end{equation}
	Equation (\ref{eq:chi_mi_formal_integration}) is then inserted into the right-hand side of Eq. (\ref{eq:chi_im_initial}) resulting in the expression 
	
	\begin{align}
			i \dfrac{\partial}{\partial t} \ket{\chi_{\vec{k}, \sigma}^{(i)}(t)}  =&  \hat{\vec{A}}^{(\vec{k}, \sigma)}_Q(t) \cdot  \bigg[ \vec{j}_{i,i}(t) \ket{\chi_{\vec{k}, \sigma}^{(i)}(t)} \nonumber \\
			  -i \sum_{m\neq i} \vec{j}_{i,m}(t) &\int_{t_i}^{t} dt' \hat{\vec{A}}^{(\vec{k}, \sigma)}_Q(t') \cdot  \vec{j}_{m,i}(t') \ket{\chi_{\vec{k}, \sigma}^{(i)}(t')} \bigg]. \label{eq:chi_i_just_before_Markov}
	\end{align}
	Following Ref. \cite{Stammer2024_MarkovPaper}, we now employ a Markov-type approximation \cite{Mandel_Wolf_1995} by expanding the state $\ket{\chi_{\vec{k}, \sigma}^{(i)}(t')}$ around $t'=t$ as
	\begin{align}
		\ket{\chi_{\vec{k}, \sigma}^{(i)}(t')}  \simeq& 	\ket{\chi_{\vec{k}, \sigma}^{(i)}(t)}  + (t'-t) \dfrac{\partial}{\partial t} 	\ket{\chi_{\vec{k}, \sigma}^{(i)}(t')} \bigg \vert_{t'=t} \nonumber \\
		& + \dfrac{(t'-t)^2}{2!} \dfrac{\partial^2}{\partial t^2} 	\ket{\chi_{\vec{k}, \sigma}^{(i)}(t')} \bigg \vert_{t'=t} + \cdots. \label{eq:chi_beyond_markov_expansion}
	\end{align}
	and only keeping the first term, i.e., letting the state become local in time and thus without memory. With this approximation, we bring Eq. (\ref{eq:chi_i_just_before_Markov}) into the expression
	\begin{align}
		i \dfrac{\partial}{\partial t} \ket{\chi_{\vec{k}, \sigma}^{(i)}(t)}  &= \hat{\vec{A}}^{(\vec{k}, \sigma)}_Q(t) \cdot   \bigg[  \vec{j}_{i,i}(t)  \nonumber \\
		  -i  \sum_{m\neq i} \vec{j}_{i,m}(t) &\int_{t_i}^{t} dt' \hat{\vec{A}}^{(\vec{k}, \sigma)}_Q(t') \cdot  \vec{j}_{m,i}(t') \bigg] \ket{\chi_{\vec{k}, \sigma}^{(i)}(t)}. \label{eq:chi_Markov_approx}
	\end{align}
In contrast to Eqs. (\ref{eq:chi_mn_decoupled}) and (\ref{eq:chi_im}) which requires the integration of $M$ photonic states, $\{ \ket{\chi^{(m)}_{\vec{k}, \sigma}(t)} \}$, Eq. (\ref{eq:chi_Markov_approx}) only requires the integration of a single state, as we only keep the state with the index $m=i$, which eases the numerical effort.

This Markov-type approximation is quantified by considering higher-order terms in the expansion in Eq. (\ref{eq:chi_beyond_markov_expansion}). Including the first-order term in $(t'-t)$ yields terms on the form
\begin{equation}
	\hat{a}_{\vec{k}, \sigma} \sum_{m \neq i} \vec{j}_{i,m}(t) \int_{t_i}^t dt' e^{-i \omega_k t'} \vec{j}_{m,i}(t') (t'-t) \times \partial_t  \ket{\chi_{\vec{k}, \sigma}^{(i)}(t)}, \label{eq:markov_type_first_order_term}
\end{equation}
with similar expressions proportional to $\hat{a}_{\vec{k}, \sigma}^\dagger$. Though the integral in Eq. (\ref{eq:markov_type_first_order_term}) in general yields a larger value than the integral in the second line of Eq. (\ref{eq:chi_Markov_approx}), the fact that Eq. (\ref{eq:markov_type_first_order_term}) is proportional to $\partial_t   \ket{\chi_{\vec{k}, \sigma}^{(i)}(t)} \propto g_0$ [see Eq. (\ref{eq:chi_im_initial})] means that the first term in Eq. (\ref{eq:chi_beyond_markov_expansion}) dominates all higher-order terms for reasonable pulse lengths. However, for sufficiently long pulses the integrals related to the higher-order terms will be on the order of $1/g_0$ such that the approximation is no longer valid. In the present work, we do not consider such long pulses.

One could consider including the higher-order terms in the expansion in Eq. (\ref{eq:chi_beyond_markov_expansion}) for improved accuracy. It turns out, however, that all higher-order terms contribute to the same order in $g_0$ and the expression cannot be truncated at a given order of $g_0$ as an infinite number of terms would have to be included. See App. \ref{App:Beyond_Markov} for more details on going beyond the first term in Eq. (\ref{eq:chi_beyond_markov_expansion}).

	\subsection{Neglecting higher-order commutators: Markov-state approximation (MSA)}
	
	 As we are interested in the origin of the nonclassical features in HHG, we manipulate Eq. (\ref{eq:chi_Markov_approx}) further for better analytical insights. Following Ref. \cite{Stammer2024_MarkovPaper}, we first use that $\sum_{m\neq i} \ket{\phi_m}\bra{\phi_m} = \mathbb{1} - \ket{\phi_i}\bra{\phi_i}$ and define 
	\begin{equation}
		\hat{W}_{\vec{k}, \sigma}(t) = \int_{t_i}^t dt' \hat{\vec{A}}^{(\vec{k}, \sigma)}_Q(t') [ \hat{\vec{j}}_H(t') - \langle \hat{\vec{j}}_H(t') \rangle], 
	\end{equation}
	where $\hat{\vec{j}}_H(t) = \hat{\mathcal{U}}^\dagger_{sc}(t) \hat{\vec{j}}(t) \hat{\mathcal{U}}_{sc}(t)$ is a Heisenberg-type formulation of the time-dependent current operator with $\hat{\mathcal{U}}_{sc}(t)$ being the time-evolution operator associated with the semiclassical Hamiltonian. Note that $\langle \hat{\vec{j}}_H(t) \rangle = \bra{\phi_i} \hat{\mathcal{U}}^\dagger_{sc}(t) \hat{\vec{j}}(t) \hat{\mathcal{U}}_{sc}(t) \ket{\phi_i} = \vec{j}_{i,i}(t)$. Equation (\ref{eq:chi_Markov_approx}) is then rewritten as
	\begin{align}
		i \dfrac{\partial}{\partial t} \ket{\chi_{\vec{k}, \sigma}^{(i)}(t)}  =& \big[\hat{\vec{A}}^{(\vec{k}, \sigma)}_Q(t) \cdot \vec{j}_{i,i}(t)  \nonumber \\
		&- i \langle \dot{\hat{W}}_{\vec{k}, \sigma}(t)\hat{W}_{\vec{k}, \sigma}(t) \rangle_{el} \big]  \ket{\chi_{\vec{k}, \sigma}^{(i)}(t)}, \label{eq:chi_Markov_W}
	\end{align}
	where $\langle \cdot \rangle_{el}$ denotes the expectation value of only the electronic operators. Equation (\ref{eq:chi_Markov_W}) yields the solution 
	\begin{equation}
		\ket{\chi_{\vec{k}, \sigma}^{(i)}(t)} =  \hat{\mathcal{D}}[\beta_{i,i}^{(\vec{k}, \sigma)}(t) ] e^{-\frac{1}{2} \langle \hat{W}_{\vec{k}, \sigma}^2(t) \rangle_{el}} \ket{0}, \label{eq:chi_Markov_solution}
	\end{equation}
	where the time-dependent displacement amplitude in the $\vec{k}$'th mode is given by
	\begin{equation}
		\beta_{m,n}^{(\vec{k}, \sigma)}(t) = -i \frac{g_0}{\sqrt{\omega_{k}}} \int_{t_i}^t dt'~ \vec{j}_{m,n}(t') \cdot \hat{\vec{e}}_\sigma^* e^{i \omega_k t'}. \label{eq:beta_general_expression}
	\end{equation}
	We will refer to Eq. (\ref{eq:chi_Markov_solution}) as the MSA as it is achieved from the Markov-type approximation in Eq. (\ref{eq:chi_Markov_approx}).
	In obtaining Eq. (\ref{eq:chi_Markov_solution}), we neglected the commutator $[\dot{\hat{W}}_{\vec{k}, \sigma}(t), \hat{W}_{\vec{k}, \sigma}(t)]$ which expresses the fluctuations of the fluctuations of the current operator. Further, only terms up to second order in $g_0$ are kept. A full derivation of Eq. (\ref{eq:chi_Markov_solution}) can be found in App. \ref{App:Derivation_of_state} and a derivation without decoupling but with an SFA approach can be found in Ref. \cite{Stammer2024_MarkovPaper}.

	We now investigate $\langle \hat{W}^2_{\vec{k}, \sigma} (t) \rangle_{el}$. It is seen from Eq. (\ref{eq:chi_Markov_solution}) that for  $\langle \hat{W}_{\vec{k}, \sigma}^2(t) \rangle_{el} = 0$ the photonic state would be a coherent state and hence have $Q_{\vec{k}, \sigma} = \eta_{\vec{k}, \sigma} = 0$. This implies that $\langle \hat{W}_{\vec{k}, \sigma}^2(t) \rangle_{el}$ is the cause of the quantum properties of the emitted HHG in the present MSA limit. We thus write out  $\langle \hat{W}_{\vec{k}, \sigma}^2(t) \rangle_{el}$ explicitly	 
	 \begin{align}
	 	\langle \hat{W}^2_{\vec{k}, \sigma}(t) \rangle_{el} =&  \int_{t_i}^t dt' \int_{t_i}^t dt'' \hat{\vec{A}}^{(\vec{k}, \sigma)}_Q(t') \hat{\vec{A}}^{(\vec{k}, \sigma)}_Q(t'')   \nonumber \\
	 	 &\times \big[ \langle  \hat{\vec{j}}_H(t')  \hat{\vec{j}}_H(t'')  \rangle-  \langle \hat{\vec{j}}_H(t')  \rangle  \langle \hat{\vec{j}}_H(t'')  \rangle  \big]. \label{eq:W_squared_JH}
	 \end{align}
	Interestingly, similar expressions with the time correlation of the current are obtained when using Heisenberg equations of motion for the photonic operators \cite{Sundaram1990}. The time-correlation function in Eq. (\ref{eq:W_squared_JH}) is numerically complicated to calculate. Rewriting, we instead express the correlations of the current in terms of the transition current matrix elements which is equally exact. To this end, we use $\mathbb{1} = \Sigma_m \ket{\phi_m}\bra{\phi_m}$ and obtain the relation

	\begin{align}
		 \langle  \hat{\vec{j}}_H(t')  \hat{\vec{j}}_H(t'')  \rangle-  \langle \hat{\vec{j}}_H(t')  \rangle  \langle \hat{\vec{j}}_H(t'')  \rangle  = \sum_{m\neq i} \vec{j}_{i,m}(t') \vec{j}_{m,i}(t''). \label{eq:fluctuations_as_transition_elements}
	\end{align}
	Equation (\ref{eq:fluctuations_as_transition_elements}) highlights an interesting equivalence between the time correlations of the (Heisenberg type) current and the transitions current matrix elements. This equation shows that the role of the transition current elements can be reformulated as time correlations of the current within the presented scheme of approximations. 
	Inserting Eq. (\ref{eq:fluctuations_as_transition_elements}) into Eq. (\ref{eq:W_squared_JH}), we find
	
	\begin{align}
		\langle \hat{W}_{\vec{k}, \sigma}^2(t) \rangle_{el} =& \sum_{m\neq i} \big[  \int_{t_i}^t dt' \hat{\vec{A}}_Q^{(\vec{k}, \sigma)}(t') \vec{j}_{i,m}(t') \big] \nonumber \\
		& \times \big[  \int_{t_i}^t dt'' \hat{\vec{A}}_Q^{(\vec{k}, \sigma)}(t'')  \vec{j}_{m,i}(t'') \big]. \label{eq:W_intermediate}
	\end{align}
	As we are interested in the final photon state after the end of the driving pulse, we let $t \rightarrow \infty$. Writing out Eq. (\ref{eq:W_intermediate}) then yields

	\begin{align}
			\langle \hat{W}_{\vec{k}, \sigma}^2 \rangle_{el} =&  B_{\vec{k}, \sigma} ~ \big [\hat{a}_{\vec{k}, \sigma}^2  e^{-i \varphi_k} + \big(\hat{a}_{\vec{k}, \sigma}^\dagger \big)^2 e^{i \varphi_k} \big ] \nonumber \\
			& +  C_{\vec{k}, \sigma} ~ \hat{a}_{\vec{k}, \sigma}^\dagger \hat{a}_{\vec{k}, \sigma} +  D_{\vec{k}, \sigma} ~ \hat{a}_{\vec{k}, \sigma} \hat{a}_{\vec{k}, \sigma}^\dagger ,
	\end{align}

	where we have defined
	\begin{subequations}\label{eq:operator_factors}
			\begin{align} 
			B_{\vec{k}, \sigma} &=  \bigg \lvert \sum_{m\neq i}  J_{i,m}^{(\vec{k}, \sigma, +)} J_{m,i }^{(\vec{k}, \sigma, +)} \bigg \rvert, \\
			\varphi_{\vec{k}, \sigma}&= \arg \bigg[\sum_{m\neq i}  J_{i,m}^{(\vec{k}, \sigma, +)} J_{m,i }^{(\vec{k}, \sigma, +)} \bigg], \\
			C_{\vec{k}, \sigma} &=  \sum_{m \neq i} \lvert  J_{i,m}^{(\vec{k}, \sigma, +)} \rvert^2, \\
			D_{\vec{k}, \sigma} &= \sum_{m \neq i} \lvert  J_{i,m}^{(\vec{k}, \sigma, -)} \rvert^2,
		\end{align}
	\end{subequations}
with the integrated transition current elements
	\begin{equation}
		 J_{m,n}^{(\vec{k}, \sigma, \pm)} =  \dfrac{g_0}{\sqrt{\omega_k}}\int_{t_i}^\infty dt' e^{\pm i \omega_k t'} \vec{j}_{m,n}(t') \cdot \hat{\vec{e}}_\sigma^{(*)}, \label{eq:J_mn_definition}
	\end{equation}
	where the polarization unit vector is complex conjugated for the positive phase. Note that $ \big[J_{m,n}^{(\vec{k}, \sigma, -)}(t)\big]^* =  J_{n,m}^{(\vec{k}, \sigma, +)}(t)$.
However, different from the case presented in Ref. \cite{Stammer2024_MarkovPaper}, we do not make further assumptions on the elements in Eq. (\ref{eq:operator_factors}).

We now calculate expectation values for the state in Eq. (\ref{eq:chi_Markov_solution}) by expanding the exponential function in Eq. (\ref{eq:chi_Markov_solution}) to second order in $g_0$, i.e., $e^{-\frac{1}{2} \langle \hat{W}^2\rangle_{el}} \simeq 1 - \frac{1}{2}  \langle \hat{W}^2\rangle_{el}$, which yields the approximate state for a given mode
\begin{equation}
	\ket{\chi_{\vec{k}, \sigma}^{(i)}} \simeq \hat{\mathcal{D}}[\beta_{i,i}^{(\vec{k}, \sigma)}(t) ] \bigg [ (1 - \dfrac{1}{2} D_{\vec{k}, \sigma} ) \ket{0} - \dfrac{1}{\sqrt{2}} B_{\vec{k}, \sigma} e^{i \varphi_{\vec{k}, \sigma}} \ket{2} \bigg]. \label{eq:chi_markov_state_approximate}
\end{equation}
The spectrum [Eq. (\ref{eq:spectrum_qo})] is then to lowest order in $g_0$ given as
\begin{align}
	S(\omega_k)& = \dfrac{\omega_k^3}{g_0^2 (2\pi)^2 c^3} \sum_{\sigma} \lvert \beta_{i,i}^{(\vec{k}, \sigma)} \rvert^2  \nonumber \\
	&= \dfrac{\omega_k^2}{(2\pi)^2 c^3} \sum_\sigma \lvert \hat{\vec{e}}_\sigma^* \cdot \tilde{\vec{j}}_{i,i}(\omega_k) \rvert^2, \label{eq:spectrum_markov_state}
\end{align}
where the definition of $\beta_{i,i}^{(\vec{k}, \sigma)}$ in Eq. (\ref{eq:beta_general_expression}) was used. We note that Eq. (\ref{eq:spectrum_markov_state}) up to constants is identical to the semiclassical spectrum in Eq. (\ref{eq:spectrum_sc}). This shows that in the limit of a weak coupling (here $g_0 = 4 \times 10^{-8}$ a.u.) the spectrum itself is dominated by the classical current, highlighting that the HHG spectrum is not a suitable observable for inferring nonclassical properties of the generated light. To investigate the nonclassical properties of light, we calculate the Mandel-Q and the squeezing parameters.
Within the MSA, the Mandel-Q parameter [Eq. (\ref{eq:mandel_Q_def})] is given by
\begin{align}
	Q_{\vec{k}, \sigma} =& \dfrac{B_{\vec{k}, \sigma}^2 + \lvert \beta_{i,i}^{(\vec{k}, \sigma)} \rvert^4 - 2  B_{\vec{k}, \sigma} \text{Re} \big[  \big(\beta_{i,i}^{(\vec{k}, \sigma)}\big)^2 e^{i \varphi_{\vec{k}, \sigma}} \big] }{ \lvert \beta_{i,i}^{(\vec{k}, \sigma)} \rvert^2 (1- D_{\vec{k}, \sigma}) + B_{\vec{k}, \sigma}^2 } \nonumber \\
	&  - \lvert \beta_{i,i}^{(\vec{k}, \sigma)} \rvert^2, \label{eq:mandel_q_markov_state}
\end{align}
where terms of order $\mathcal{O}(g_0^4)$ have been included for the necessary numerical stability. The quadrature variance minimized to calculate the squeezing [Eq. (\ref{eq:squeezing_def})] is within the MSA given by
\begin{equation}
	 [\Delta \hat{X}_{\vec{k}, \sigma}(\theta)]^2 = \dfrac{1}{4} \big[1 - 2 B_{\vec{k}, \sigma}  \cos(2 \theta - \varphi_{\vec{k}, \sigma}) \big]. \label{eq:squeezing_markov_state}
\end{equation}
We note, that the quantum features calculated by Eqs. (\ref{eq:mandel_q_markov_state}) and (\ref{eq:squeezing_markov_state}) are dependent on the transition currents [or equivalently the current correlations, see Eq. (\ref{eq:fluctuations_as_transition_elements})] that in general have different spectral features than the classical current. This means that the quantum features do not follow the spectral structure, e.g., with peaks placed at odd harmonics, and consequently will not have the same selection rules. 
 
\begin{figure}
	\centering
	\includegraphics[width=1\linewidth]{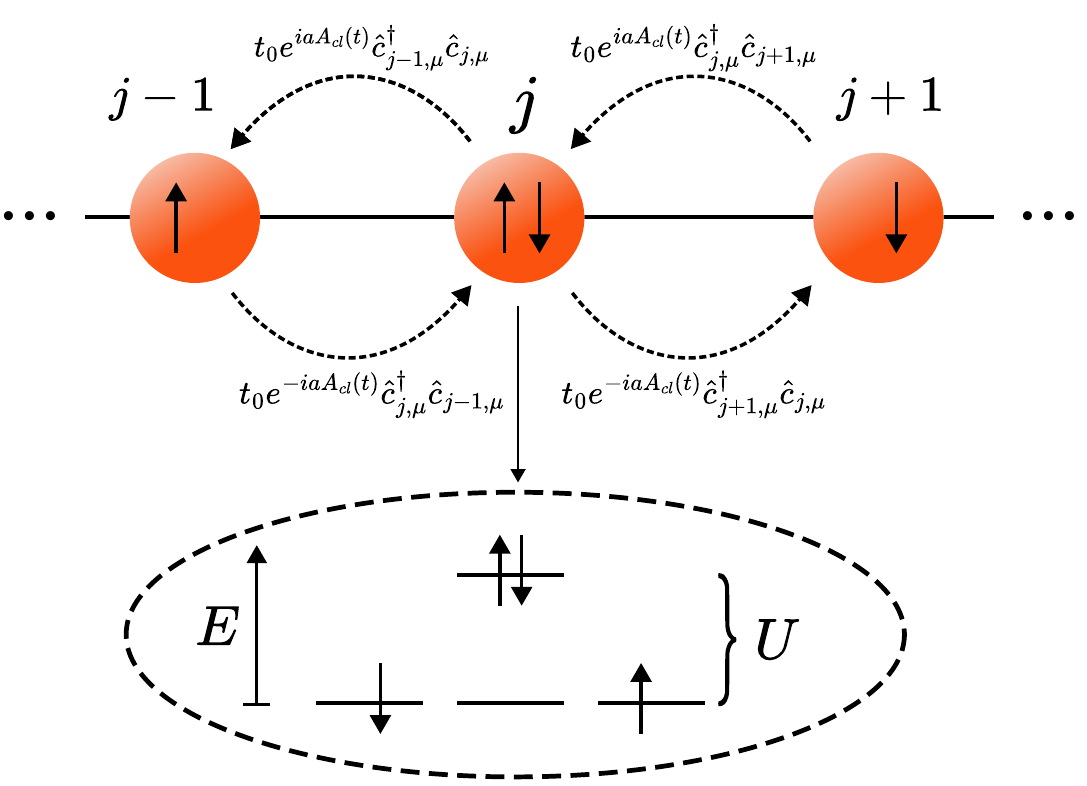}
	\caption{Illustration of the Fermi-Hubbard model. The top figure shows the nearest-neighbor hopping while the bottom diagram shows the energy cost of $U$ associated with a single doubly-occupied site, a doublon.}
	\label{fig:newfhmodel}
\end{figure}

We further emphasize that Eqs. (\ref{eq:spectrum_markov_state})-(\ref{eq:squeezing_markov_state}) only require solving the TDSE of the electronic system driven by a classical field [Eq. (\ref{eq:TDSE_SC})] and do not require the integration of any additional equations of motion for the photonic state as is necessary for the previous levels of approximations [Eqs. (\ref{eq:chi_mn_decoupled}, \ref{eq:chi_im}, \ref{eq:chi_Markov_approx})], easing the numerical effort significantly. 

Finally, we note that if one uses the MSA without the decoupling of modes, one would end up with the same expressions for the examined observables as in Eqs. (\ref{eq:spectrum_markov_state}, \ref{eq:mandel_q_markov_state}, \ref{eq:squeezing_markov_state}). This shows that the single-mode observables as considered here are not affected by the decoupling of photonic modes in the present MSA. Consequently, the errors introduced when deriving the MSA are larger than the errors introduced by decoupling the photonic modes. Unfortunately, it is not possible to test the case of coupled photonic modes in the lower levels of approximations (see Fig. \ref{fig:hierachyofapproximations}) due to numerical constraints, and as such it is challenging to quantify the validity of the photonic decoupling. 
Studying these levels of approximations in a simpler electronic system in an appropriate limit where coupled photonic modes can be considered and by extension also considering multi-mode observables would deserve an entire manuscript on its own.

 \section{Electronic system} \label{sec:electronic_system}
 
  \begin{figure*}
 	\begin{center}
 		\includegraphics[width=18cm]{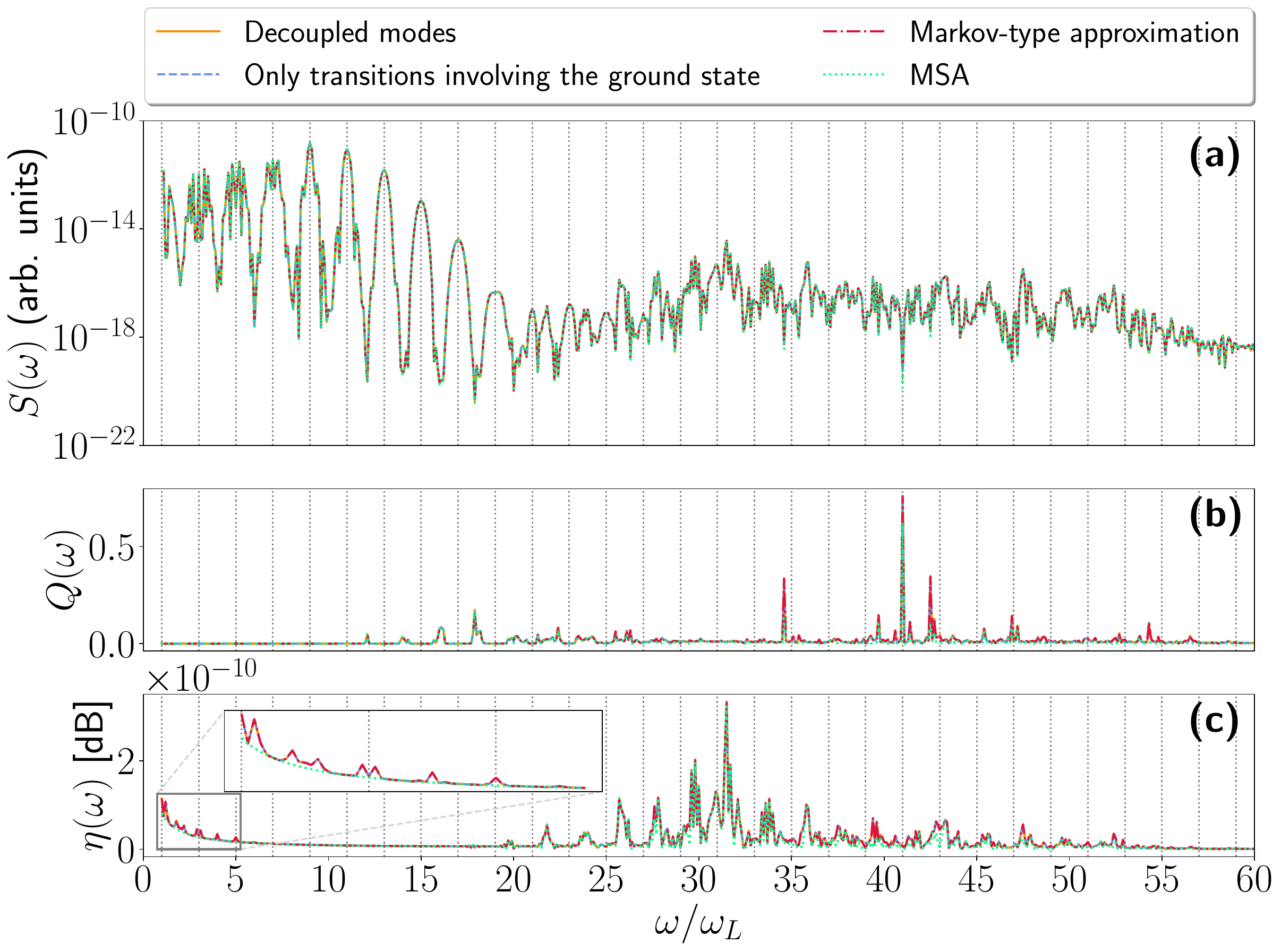}
 	\end{center}
 	\caption{Results for the various degrees of approximation [see Fig. \ref{fig:hierachyofapproximations}] for an $N_c = 10$ cycle pulse, see text for additional parameters. (a) The spectrum [Eq. (\ref{eq:spectrum_qo})]. We note that no level of approximation shows any significant difference. (b) The Mandel-Q parameter [Eq. (\ref{eq:mandel_Q_def})], and (c) the squeezing [Eq. (\ref{eq:squeezing_def})]. We note that the calculation in the decoupled limit [Eq. (\ref{eq:chi_mn_decoupled}), orange full line], only including transitions involving the electronic ground state [Eq. (\ref{eq:chi_im}), blue dashed line], and the Markov-type approximation [Eq. (\ref{eq:chi_Markov_approx}), red dashdotted line] match very well for all frequencies. The MSA [Eq. (\ref{eq:chi_markov_state_approximate}), green dotted line] also shows good agreement with all results but has some deviations at various harmonics in both the Mandel-Q parameter and squeezing. We also note that the MSA does not capture the squeezing at the lower harmonics as seen in the insert in (c).}
 	\label{fig:methodcomparisonfulldatal8u10nc10}
 \end{figure*}

 The derivations in Secs. \ref{sec:theory} and \ref{sec:hierarchy_of_approx} are completely general for any electronic system one might consider. In the present paper, we use the driven Fermi-Hubbard model as the electronic system. This model was recently shown to yield nonclassical harmonics in the so-called Mott-insulating limit \cite{Lange2024_electroncorrelation}. That is, we take the semiclassical Hamiltonian to be that of the field-driven Fermi-Hubbard model, i.e., $\hat{H}_{sc}(t) \rightarrow \hat{H}_{FH}(t)$. The Fermi-Hubbard model is a generic many-body model that captures beyond-mean-field electronic interaction with the Hubbard-$U$. This choice of model allows us to simulate the semiclassical TDSE without any further approximations as all states can be included in the numerical modeling for a small enough chain. We specifically consider a model with periodic boundary conditions at half filling with an equal number of spin-up and spin-down electrons. Within the dipole approximation, the system driven by a classical laser pulse is described by the time-dependent Hamiltonian \cite{Essler_Hubbard_book}
 \begin{equation}
 	\hat{H}_{sc}(t) = \hat{H}_{FH}(t) = \hat{H}_{hop}(t) + \hat{H}_U, \label{eq:H_FH}
 \end{equation}
 with
 \begin{align}
 	\hat{H}_{hop}(t) &= - t_0 \sum_{j, \mu} \big( e^{ia A_{cl}(t)} \hat{c}_{j, \mu}^\dagger \hat{c}_{j+1, \mu} + \text{H.c.} \big), \nonumber \\
 	\hat{H}_U &= U \sum_j (\hat{c}_{j, \uparrow}^\dagger \hat{c}_{j, \uparrow}) (\hat{c}_{j, \downarrow}^\dagger \hat{c}_{j, \downarrow}),
 \end{align}
 where $t_0$ is the hopping matrix element for an electron to hop from site $j$ to site $j \pm 1$, the operator $\hat{c}^\dagger_{j,\mu}$ ($\hat{c}_{j,\mu}$) creates (annihilates) an electron with spin $\mu \in \{ \uparrow, \downarrow\} $ on site $j$, $a$ is the lattice constant and $U$ describes the degree of beyond-mean-field onsite electron-electron repulsion. We only include nearest-neighbor hopping which is the common limit of this model \cite{Hansen22, Hansen22_2, Murakami2021, Lange2024_electroncorrelation, Hansen_2024_lattice_imperfections}. A schematic overview of the model is given in Fig. \ref{fig:newfhmodel}. In this model, the time-dependent current operator is given as \cite{Lange2024_electroncorrelation,  Essler_Hubbard_book}

 \begin{figure*}
 	\begin{center}
 		\includegraphics[width=18cm]{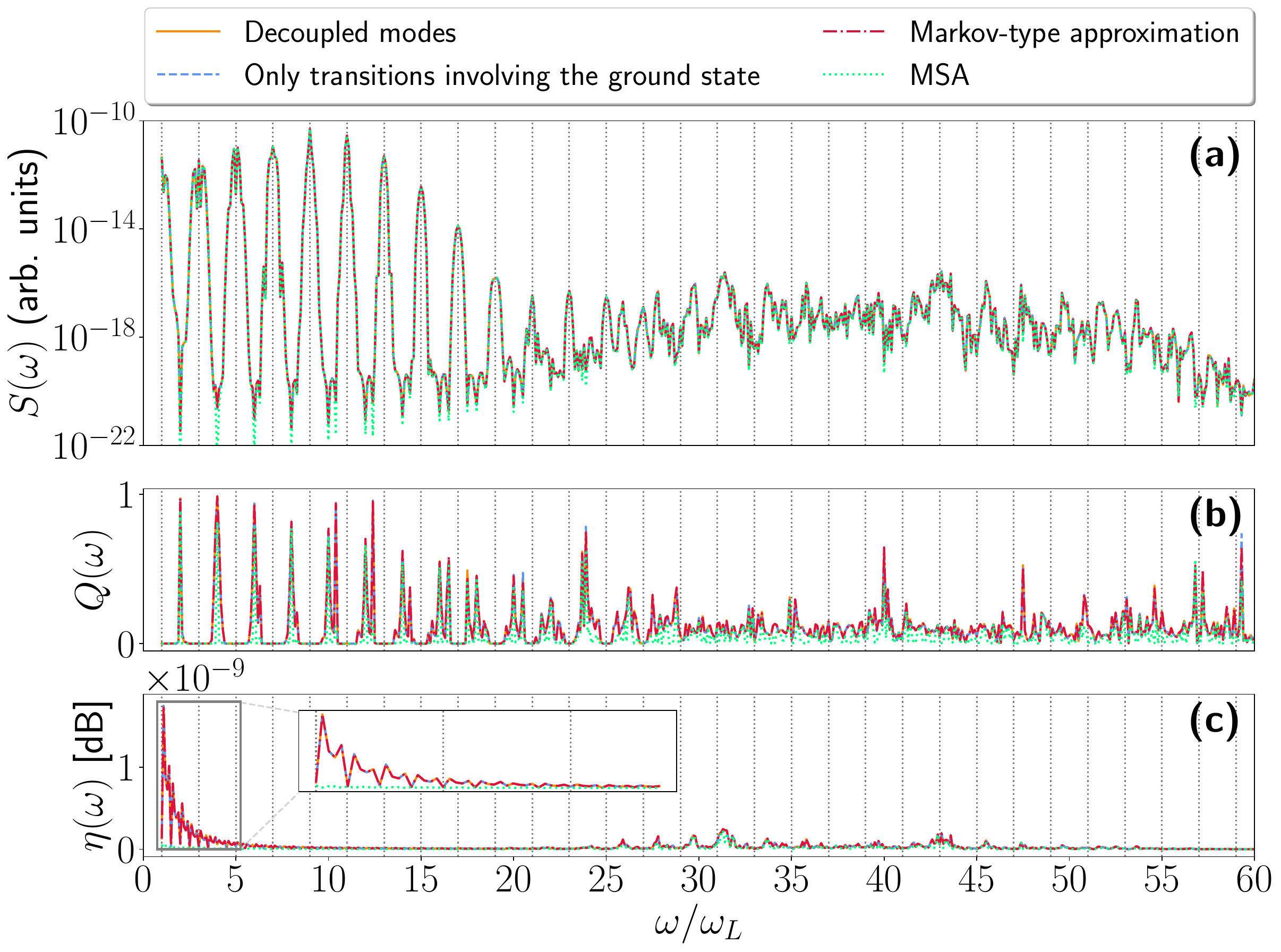}
 	\end{center}
 	\caption{Results for the various degrees of approximation [see Fig. \ref{fig:hierachyofapproximations}] for an $N_c = 18$ cycle pulse, see text for additional parameters. (a) The spectrum [Eq. (\ref{eq:spectrum_qo})]. We note that no level of approximation shows any significant deviation from the others. (b) The Mandel-Q parameter [Eq. (\ref{eq:mandel_Q_def})]. We note that the calculation in the decoupled limit [Eq. (\ref{eq:chi_mn_decoupled}), solid orange], the limit that only includes transitions involving the electronic ground state [Eq. (\ref{eq:chi_im}), dashed blue] as well as the Markov-type approximation [Eq. (\ref{eq:chi_Markov_approx}), dashdotted red] match well for all frequencies. The MSA [Eq. (\ref{eq:chi_markov_state_approximate}), dotted green] matches the more exact results well, but deviates more for higher harmonics. (c) The squeezing [Eq. (\ref{eq:squeezing_def})]. Here all levels of approximations match well, except for the MSA which does not capture the squeezing at lower harmonics captured by the other methods as seen in the inset.} 
 	\label{fig:methodcomparisonfulldatal8u10nc18}
 \end{figure*}
 \begin{equation}
 	\hat{\vec{j}}(t) = - i a t_0  \sum_{j, \mu} \big( e^{ia A_{cl}(t)} \hat{c}_{j, \mu}^\dagger \hat{c}_{j+1, \mu} - \text{H.c.} \big) \hat{\vec{x}}, \nonumber \\
 \end{equation}
 which is in the direction of the Fermi-Hubbard chain, taken to be in the $\hat{\vec{x}}$-direction. For details on the model see, e.g., Ref. \cite{Lange2024_electroncorrelation, Essler_Hubbard_book}.
 
 In this work, we use a chain of $L=8$ sites (with periodic boundary conditions), a lattice spacing of $a = 7.5589$ a.u., and $t_0 = 0.0191$ a.u. picked specifically to match those of the cuprate Sr$_2$CuO$_3$ \cite{Tomita2001} as done previously in Refs. \cite{Silva2018, Hansen22, Hansen22_2, Lange2024_electroncorrelation, Lange2024_noninteger, Hansen_2024_lattice_imperfections}. We investigate the Mott-insulating phase with $U= 10 t_0$ which highly favors anti-ferromagnetic ordering in the field-free ground state of the system. For the driving field, we use a $\sin^2$ envelope function for a pulse polarized along the lattice dimension with $N_c$ cycles
 
 \begin{equation}
 	A_{cl}(t) = A_0 \sin(\omega_L t + \pi/2) \sin^2 \bigg(\dfrac{\omega_L t}{2 N_c} \bigg),
 \end{equation}
 where the laser frequency is $\omega_L = 0.005$ a.u. $=33$ THz, the vector potential is chosen to be $A_0 = F_0/\omega_L = 0.194$ a.u. This corresponds to a peak intensity of $3.3 \times 10^{10}$ W/cm$^2$, well below the expected damage threshold. In order to obtain the transition current elements $\vec{j}_{m,n}(t)$, we solve Eq. (\ref{eq:TDSE_SC}) for all states $\{\ket{\phi_m(t)}\}$ using the Arnoldi-Lancoz algorithm \cite{Park1986, Smyth1998, Guan2007, Frapiccini2014} with a Krylov subspace of dimension $4$. We use a time step of $\Delta t= 1/\sqrt{10}$ a.u. and all results have been checked for convergence. To limit the dimensionality of the required Hilbert space, we start from a spin-symmetric ground state with vanishing total crystal momentum. As the Hamiltonian Eq.(\ref{eq:H_FH}) is invariant under spin-flip and conserves total crystal momentum only states within that subspace are needed.

\section{Results and discussion} \label{sec:results_and_discussion}
We calculate both the HHG spectrum, photon statistics, and squeezing for all levels of approximations. When solving the photonic equations of motion for a given level of approximation [Eqs. (\ref{eq:chi_mn_decoupled}), (\ref{eq:chi_im}), and (\ref{eq:chi_Markov_approx})], the photonic states are expanded in a Fock-basis truncated at $50$ photons per mode and have been tested for convergence by truncating at $100$ photons. After the end of the driving pulse, we rarely find population in a Fock state with more than two photons. Details on the calculation of expectation values can be found in Ref. \cite{Lange2024_electroncorrelation}. 

The results for a simulation with a laser pulse of $N_c = 10$ cycles are shown in Fig. \ref{fig:methodcomparisonfulldatal8u10nc10}. Looking at the spectrum [Fig. \ref{fig:methodcomparisonfulldatal8u10nc10}(a)], we note two different regimes. At lower harmonics ($\omega/\omega_L \leq 19$), we see regular peaks in the signal at odd harmonics. This part of the spectrum is due to the so-called intrasubband current. At higher harmonics ($\omega/\omega_L > 19$), the spectrum is generated by the so-called intersubband current. We note that the signal in this part of the spectrum is more irregular with peaks at non-integer harmonics. The origin of the presence of this signal at non-integer harmonics is due to a population of several Floquet states. A more detailed discussion of the spectrum can be found in Refs. \cite{Lange2024_electroncorrelation, Lange2024_noninteger}. Comparing the different levels of approximation in Fig. \ref{fig:methodcomparisonfulldatal8u10nc10}(a), we find that all levels of approximation produce the same spectrum with no noticeable difference. This highlights that in the regime of weak coupling ($g_0 = 4 \times 10^{-8}$ a.u.), the quantum optical nature of HHG does not appear in the spectrum. Looking at the photon statistics calculated via the Mandel-Q parameter [Fig. \ref{fig:methodcomparisonfulldatal8u10nc10}(b)], we first note that finite values are only seen at distinct typically non-odd harmonic frequencies, clearly showing that the Mandel-Q parameter does not peak at the same frequencies as the spectrum in Fig. \ref{fig:methodcomparisonfulldatal8u10nc10}(a). This can be understood from the MSA [Eq. (\ref{eq:chi_Markov_solution})] where it is seen that all the transition currents, $\vec{j}_{i,m}(t)$, are included to calculate the Mandel-Q parameter [see Eqs. (\ref{eq:operator_factors}, \ref{eq:mandel_q_markov_state})]. These transition current elements have different spectral features than the diagonal current, $\vec{j}_{i,i}(t)$, used to generate the spectrum \cite{Lange2024_electroncorrelation}, which is why the Mandel-Q parameter peaks at other frequencies. Looking at Fig. \ref{fig:methodcomparisonfulldatal8u10nc10}(b), we see that all levels of approximations qualitatively show the same features. The MSA [Eq. (\ref{eq:chi_markov_state_approximate}), dotted green], however, shows some relatively minor deviations at certain harmonics. For the squeezing [Fig. \ref{fig:methodcomparisonfulldatal8u10nc10}(c)] we see two spectral regions with nonvanishing squeezing. At the lowest harmonics ($\omega/\omega_L \leq 5$) we see a small degree of squeezing. Most squeezing is, however, seen in the intersubband region ($\omega/\omega_L > 19$). As for the Mandel-Q parameter, we note that the degree of squeezing does not peak at the same harmonics as the spectrum in Fig. \ref{fig:methodcomparisonfulldatal8u10nc10}(a). This can again be understood from the MSA, where we see that the quantities involved in calculating the degree of squeezing [Eq. (\ref{eq:squeezing_markov_state})] contain contributions from the transition current elements with different spectral features than those in the classical current. Comparing the different levels of approximations, we also find good agreement between the produced results for the degree of squeezing. Note that the MSA does not capture the squeezing at the lower harmonics as shown in the inset in Fig. \ref{fig:methodcomparisonfulldatal8u10nc10}(c). This squeezing results from higher-order terms not included in the derivation of Eq. (\ref{eq:chi_markov_state_approximate}). 
\begin{figure}
	\centering
	\includegraphics[width=1\linewidth]{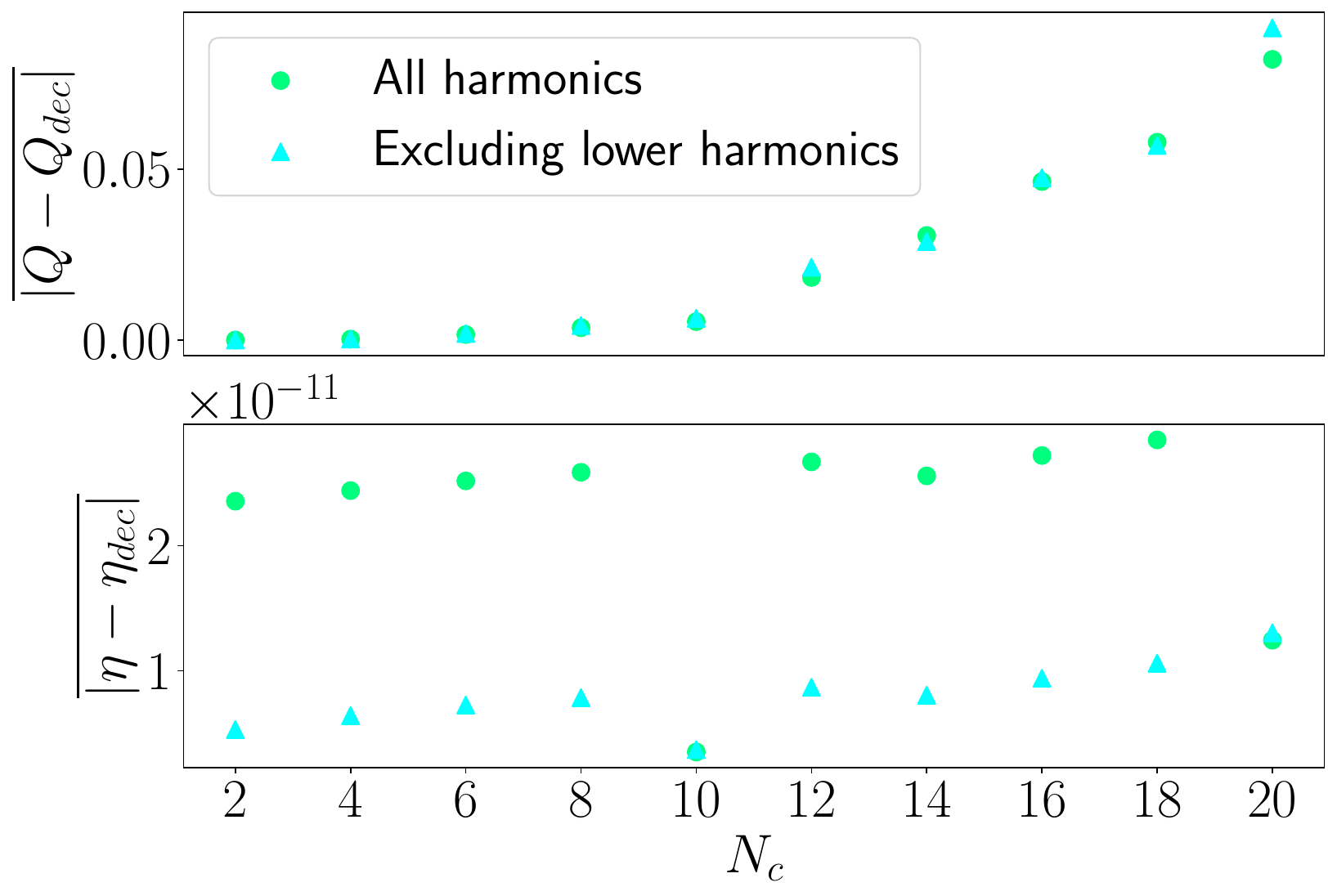}
	\caption{The absolute difference of the Mandel-$Q$ parameter (upper panel) and squeezing parameter (lower panel) between the exact solution [Eq. (\ref{eq:chi_mn_decoupled})] and the MSA averaged over all considered harmonics ($1 \leq \omega/\omega_L \leq 60$, green circles) and all higher harmonics ($10 \leq \omega/\omega_L \leq 60$, cyan triangles) for various number of pulse lengths determined by the number of cycles, $N_c$. For the Mandel-Q parameter, we see that the deviation grows with increasing pulse length. Similarly, the squeezing deviates slightly more for increasing pulse length. We note, that for $N_c = 10, 20$, the system has less squeezing in the lower harmonics which improves the MSA as it does not capture the degree of squeezing in the lower harmonics.}
	\label{fig:quantumdifferencepulsescalingl8u10}
\end{figure}

We also consider a longer pulse of $N_c=18$ cycles shown in Fig. \ref{fig:methodcomparisonfulldatal8u10nc18} to see how the different levels of approximations match in that case. Recall that the Markov-type approximation and the MSA rely on a local-time approximation [Eqs. (\ref{eq:chi_beyond_markov_expansion}, \refeq{eq:chi_Markov_approx})] which is expected to be less accurate for a longer pulse duration. Again for this case of $N_c=18$ cycles, we see in Fig. \ref{fig:methodcomparisonfulldatal8u10nc18}, that the spectrum consists of a regular intrasubband region and a more irregular intersubband region as in Fig. \ref{fig:methodcomparisonfulldatal8u10nc10}(a). Also for the longer pulse duration of $N_c = 18$ cycles in Fig. \ref{fig:methodcomparisonfulldatal8u10nc18}(a), all levels of approximations produce the same spectrum, though the MSA shows some minor deviations on lower even harmonics. The photon statistics characterized by the Mandel-Q parameter is significantly different for the longer pulse of $N_c=18$ cycles in Fig. \ref{fig:methodcomparisonfulldatal8u10nc18}(b) than for the shorter pulse of $N_c=10$ cycles in Fig. \ref{fig:methodcomparisonfulldatal8u10nc10}(b). We see two different regions in Fig. \ref{fig:methodcomparisonfulldatal8u10nc18}(b). One region at lower harmonics ($\omega_L/\omega \leq 25)$ with sharp peaks in the signal and a region at higher harmonics ($\omega/\omega_L > 25$) with smaller and less distinct peaks. This difference can partially be explained by the MSA. By investigating the expression for the Mandel-Q parameter for the MSA [Eq. (\ref{eq:mandel_q_markov_state})], we see that the denominator is dominated by $\lvert \beta_{i,i}^{(\vec{k}, \sigma)} \rvert^2$ which is proportional to the spectrum [see Eq. (\ref{eq:spectrum_markov_state})]. Now, as the pulse duration is increased much less signal is found at even harmonics, i.e, $\lvert \beta_{i,i}^{(\vec{k}, \sigma)} \rvert^2$ is significantly smaller at even harmonics for the longer pulse, which can be seen by comparing Figs. \ref{fig:methodcomparisonfulldatal8u10nc18}(a) and \ref{fig:methodcomparisonfulldatal8u10nc10}(a). From Eq. (\ref{eq:mandel_q_markov_state}), we see that this drop in signal at even harmonics yields a larger value for the Mandel-Q parameter. Experimentally, this means that if any signal at these lower even harmonics is measured, the photon distribution will be highly nonpoissonian. Comparing the different levels of approximations, we see that they yield close to identical results for the Mandel-Q parameter. The MSA, however, does deviate for certain harmonics and is generally less accurate for higher harmonics. The squeezing parameter for the longer pulse of $N_c = 18$ cycles is shown in Fig. \ref{fig:methodcomparisonfulldatal8u10nc18}(c). Opposite to the situation for the $N_c=10$ cycle pulse in Fig. \ref{fig:methodcomparisonfulldatal8u10nc10}(c), we now see that the degree of squeezing is largest for the lower harmonic ($\omega/\omega_L \leq  5$) which is almost an order of magnitude larger than the largest value for the squeezing parameter for the $N_c=10$ cycle pulse. This difference in magnitude is due to a different response of the electronic system, and we do not believe this to be a general feature for all electronic systems. For the $N_c=18$ cycle pulse we still see a finite degree of squeezing at higher harmonics ($\omega/\omega_L >25)$ comparable to the values seen in Fig. \ref{fig:methodcomparisonfulldatal8u10nc10}(c). Comparing the different levels of approximations for the squeezing in Fig. \ref{fig:methodcomparisonfulldatal8u10nc18}(c), we see a good agreement between all produced results. Notably, however, the MSA does again not capture the degree of squeezing at lower harmonics.

The deviation of the MSA in both the Mandel-Q parameter and squeezing from the most exact obtainable solution [Eq. (\ref{eq:chi_mn_decoupled})] is shown in Fig. \ref{fig:quantumdifferencepulsescalingl8u10}. We compute the mean of this deviation over the considered harmonics ($\omega/\omega_L \leq 60$) as $\overline{|Q-Q_{dec}|}$ for different pulse lengths with similar expressions for the squeezing, $\eta$. Here, the bar denotes the average over the considered harmonics, and the subscript $dec$ refers to the solution obtained with decoupling of the harmonic modes as the only approximation [Eq. (\ref{eq:chi_mn_decoupled})]. We note that averaging the deviation across all considered harmonics only partially yields a truthful measure of the validity of the approximation as it does not capture if the deviation is a general trend or related to specific harmonics. We first note that the accuracy of the Mandel-Q parameter deviates with an increasing pulse length. Further, we see that the MSA does not exactly capture the squeezing. The offset in the deviation is due to the lower harmonics not being captured. For the present system, the squeezing at lower harmonics is less for $N_c=10$ cycles than for, e.g., $N_c=18$ cycles as seen in Fig. \ref{fig:quantumdifferencepulsescalingl8u10} and by comparing Figs. \ref{fig:methodcomparisonfulldatal8u10nc10} and \ref{fig:methodcomparisonfulldatal8u10nc18}. Further, we also note a slight increase in the deviations for the squeezing. These large deviations at lower harmonics are due to higher-order commutators that are neglected in the derivation of the MSA in Eq. (\ref{eq:chi_Markov_solution}). Figure \ref{fig:quantumdifferencepulsescalingl8u10} shows as expected that the accuracy of the MSA decreases with increasing pulse duration.

\section{Conclusion and outlook} \label{sec:conclusion_and_outlook}
In this work, we derived and verified a hierarchy of approximations on the equations of motion for the quantum state of light emitted from HHG. Each step in the hierarchy of approximations, which are general for all types of electronic systems, was tested numerically using the Fermi-Hubbard model in the Mott-insulating phase. This model allowed us to solve all semiclassical TDSEs required without any further approximations. We found that including only couplings that involve the ground state is a good approximation to the full system for all considered pulse lengths. Going further, a Markov-type approximation and an even further related approximated state, the MSA, shows good agreement with the more exact results. Notably, the spectrum calculated based on these approximations matched more exact results for all pulse lengths considered. The MSA does, however, deviate from more exact results for the Mandel-Q parameter and the squeezing with increased pulse lengths. Especially, it does not capture the squeezing of the lower harmonics.

The analytical insights based on the Markov-type approximation leading to the MSA, highlight that the nonclassical features of the light are due to time correlations of the current (or dipole for atomic systems). Further, the time-correlations of the current are related to the transition current elements which is an important connection highlighting the physical relevance of the latter. By investigating the expressions for the nonclassical features of the emitted light in the MSA, we found that these have different spectral features than the HHG spectrum, as these are calculated from the transitions current elements whose spectral features are different from that of the classical current generating the HHG spectrum. We also emphasize that the approximations in this paper assume a small coupling, $g_0$, to the quantized electromagnetic field as is the typical experimental situation. Hence, the presented equations and results would become less accurate for a larger value of $g_0$. Indeed it would be worth pursuing experimental setups that would enhance the coupling to the quantized field, e.g., a cavity, as this would increase the nonclassical properties of the emitted HHG radiation \cite{Yi2024}. In this connection, we note that alternative approaches addressing this problem begin to appear: very recent work considers in reduced dimension a single-active electron coupled to a single quantized photon mode in a cavity by both an exact quantum electrodynamical approach and a semiclassical multi-trajectory simulation \cite{delapena2024quantumelectrodynamicshighharmonic}.  

As an outlook, the validated approximations may be an aid when considering the quantum backaction from the electronic system to the radiation field with a nonclassical driving field. In the theory for nonclassical driving \cite{Tzur2023_photon_statistics_force, Gorlach2023_HHG_driven_by, Tzur2024_generation_of_squeezed}, the TDSE needs to be integrated for many different classical driving fields. Without any approximations or limiting cases for which the photonic state can be analytically derived, a subsequent integration of the equations of motion for the photonic field is required, making it numerically demanding and tedious to consider a general electronic system with potential backaction onto the quantum field. However, with the explicit expression of the MSA, it might be feasible to ease the numerical effort such that the quantum backaction onto the quantized field can be studied with a nonclassical driving field, and as such the MSA can aid in the generation of nonclassical states of light in the XUV-region with applications in quantum information, sensing, and technology.

\begin{acknowledgments}
	We thank Thomas Hansen for numerical simulations of the classically driven Fermi-Hubbard model and Philipp Stammer and Rasmus Vesterager Gothelf for discussions. This work was supported by the Danish
	National Research Foundation through the Center of Excellence for Complex Quantum Systems (Grant Agreement No. DNRF152).
\end{acknowledgments}

	\appendix
	
	\section{Higher-order terms in the Markov-type approximation} \label{App:Beyond_Markov}
	In this appendix, we show how going beyond the Markov-type approximation presented in Eq. (\ref{eq:chi_Markov_approx}) by including higher-order terms in the expansion in Eq. (\ref{eq:chi_beyond_markov_expansion}) leads to an expression that cannot be truncated consistently in orders of $g_0$. 	Inserting the higher-order expansion [Eq. (\ref{eq:chi_beyond_markov_expansion})] into Eq. (\ref{eq:chi_i_just_before_Markov}) yields
	
	\begin{align}
		i \dfrac{\partial}{\partial t}  \ket{\chi_{\vec{k}, \sigma}^{(i)}(t)}& = \hat{\vec{A}}^{(\vec{k}, \sigma)}_Q(t) \cdot \vec{j}_{i,i}(t) \ket{\chi_{\vec{k}, \sigma}^{(i)}(t)}\nonumber \\
		&  -i  \hat{\vec{A}}^{(\vec{k}, \sigma)}_Q(t) \cdot \sum_{m\neq i} \vec{j}_{i,m}(t) \nonumber \\
		& \times \int_{t_i}^{t} dt' \hat{\vec{A}}^{(\vec{k}, \sigma)}_Q(t') \cdot  \vec{j}_{m,i}(t') \nonumber \\
		& \times \bigg[ \	\ket{\chi_{\vec{k}, \sigma}^{(i)}(t)}  + (t'-t) \dfrac{\partial}{\partial t'} 	\ket{\chi_{\vec{k}, \sigma}^{(i)}(t')} \bigg \vert_{t'=t} \nonumber \\
		& + \dfrac{(t'-t)^2}{2!} \dfrac{\partial^2}{\partial t'^2} 	\ket{\chi_{\vec{k}, \sigma}^{(i)}(t')} \bigg \vert_{t'=t} + \cdots \bigg].
	\end{align}
	Calculating the first and second derivative to lowest order in $g_0$ we find
	
	\begin{align}
		\dfrac{\partial}{\partial t'}  \ket{\chi_{\vec{k}, \sigma}^{(i)}(t')}\bigg \vert_{t'=t} \simeq &  - i  \hat{\vec{A}}^{(\vec{k}, \sigma)}_Q(t) \cdot  \vec{j}_{i,i}(t)  \ket{\chi_{\vec{k}, \sigma}^{(i)}(t)},  \label{eq:app_chi_first_deriv} \\
		\dfrac{\partial^2}{\partial t'^2}  \ket{\chi_{\vec{k}, \sigma}^{(i)}(t')}\bigg \vert_{t'=t}  \simeq& - i \bigg\{ \bigg[\dfrac{\partial}{\partial t}   \hat{\vec{A}}^{(\vec{k}, \sigma)}_Q(t) \bigg] \cdot \vec{j}_{i,i}(t) \nonumber \\
		& +  \hat{\vec{A}}^{(\vec{k}, \sigma)}_Q(t)  \dfrac{\partial}{\partial t}  \vec{j}_{i,i}(t) \bigg\}  \ket{\chi_{\vec{k}, \sigma}^{(i)}(t)}, \label{eq:app_chi_second_deriv}
	\end{align}
	All higher-order terms would then also yield terms that are linear in $\hat{\vec{A}}^{(\vec{k}, \sigma)}_Q(t)$ and hence $g_0$. However, we numerically find that the two terms in Eq. (\ref{eq:app_chi_second_deriv}) are comparable in magnitude which prevents us from consistently truncating the expansion of the state in Eq. (\ref{eq:chi_beyond_markov_expansion}) to a given order in $g_0$. Consequently, one cannot go beyond the leading order in the Markov-type approximation consistently with a finite number of terms and hence we consider only the Markov-type approximation in Eq. (\ref{eq:chi_Markov_approx}) in the present work. Of course, one could truncate the higher-order terms when the numerical value of the related integrals reaches a certain threshold. However, this procedure would rely on the specifics of the electronic system and laser parameters.

	\section{Derivation of the quantum state in Eq. (\ref{eq:chi_Markov_solution})} \label{App:Derivation_of_state}
	Here we derive the quantum state in Eq. (\ref{eq:chi_Markov_solution}) as the solution to Eq. (\ref{eq:chi_Markov_W}). We do this by multiplying Eq. (\ref{eq:chi_Markov_solution}) with a time-dependent phase, $e^{-i b(t)}$ and insert it into Eq. (\ref{eq:chi_Markov_W})
	
	\begin{align}
		i \dfrac{\partial}{\partial t} e^{-i b(t)} \ket{\chi_{\vec{k}, \sigma}^{(i)}(t)} =& i  \dfrac{\partial}{\partial t} e^{-i b(t)} \hat{\mathcal{D}}[\beta_{i,i}^{(\vec{k}, \sigma)}(t)] e^{-\frac{1}{2} \langle \hat{W}^2(t) \rangle_{el}} \ket{0} \nonumber \\
		=& b(t) e^{-i b(t)}  \ket{\chi_{\vec{k}, \sigma}^{(i)}(t)} \nonumber \\
		&+  i e^{-i b(t)}  \bigg[\dfrac{\partial}{\partial t} \hat{\mathcal{D}}[\beta_{i,i}^{(\vec{k}, \sigma)}(t)] \bigg] e^{-\frac{1}{2} \langle \hat{W}^2_{el}(t) \rangle} \ket{0} \nonumber \\
		& + i e^{-i b(t)} \hat{\mathcal{D}}[\beta_{i,i}^{(\vec{k}, \sigma)}(t)] \dfrac{\partial}{\partial t} e^{-\frac{1}{2} \langle \hat{W}_{el}^2(t) \rangle} \ket{0}. \label{eq:app_time_derivative_full} 
	\end{align}
	 We now calculate the time derivative of the displacement operator in the second line in Eq. (\ref{eq:app_time_derivative_full})

	\begin{align}
		&\dfrac{\partial}{\partial t} \hat{\mathcal{D}}[\beta_{i,i}^{(\vec{k}, \sigma)}(t)] = \nonumber \\
		 &\bigg\{ \dfrac{1}{2} [\dot{\beta}_{i,i}^{(\vec{k}, \sigma)*}(t) \beta_{i,i}^{(\vec{k}, \sigma)}(t) - \beta_{i,i}^{(\vec{k}, \sigma)*}(t) \dot{\beta}_{i,i}^{(\vec{k}, \sigma)}(t)]  \nonumber \\
		&+  \dot{\beta}_{i,i}^{(\vec{k}, \sigma)}(t) \hat{a}_{\vec{k}, \sigma}^\dagger - \dot{\beta}_{i,i}^{(\vec{k}, \sigma)*}(t) \hat{a}_{\vec{k}, \sigma} \bigg\} \hat{\mathcal{D}}[\beta_{i,i}^{(\vec{k}, \sigma)}(t)],
	\end{align}

	We can, upon comparison with Eq. (\ref{eq:chi_Markov_W}), deduce that the time-dependent amplitude, $\beta_{i,i}^{(\vec{k}, \sigma)}(t)$, has to be the expression given in Eq. (\ref{eq:beta_general_expression}). The first and second line in Eq. (\ref{eq:app_time_derivative_full}) thus corresponds to the first term on the right hand side of Eq. (\ref{eq:chi_Markov_W}) by absorbing the constant terms into $b(t)$. This means that the third line in Eq. (\ref{eq:app_time_derivative_full}) must yield the second term on the right hand side of Eq. (\ref{eq:chi_Markov_W}). 
	
	We now calculate the time derivative of the exponential operator in the third line in Eq. (\ref{eq:app_time_derivative_full}). A derivative of an exponential of a general operator, $\hat{F}(t)$, is given by \cite{Merzbacher_QM_book}
	\begin{align}
		\dfrac{d}{dt} e^{\hat{F}(t)} &= \big[ \dot{\hat{F}}(t) + \dfrac{1}{2!} [\hat{F}(t),\dot{\hat{F}}(t)] \nonumber \\
		&+ \dfrac{1}{3!} [\hat{F}(t), [\hat{F}(t), \dot{\hat{F}}(t)]] + \dots \big] e^{\hat{F}(t)}.
	\end{align}
	In the present case, we have to calculate 
	\begin{align}
		\dfrac{d}{dt} e^{-\frac{1}{2}  \langle \hat{W}^2_{\vec{k}, \sigma}(t) \rangle_{el}} = &- \langle  \dot{\hat{W}}_{\vec{k}, \sigma}(t)  \hat{W}_{\vec{k}, \sigma}(t) \rangle_{el} ~ e^{-\frac{1}{2} \langle  \hat{W}_{\vec{k}, \sigma}^2(t) \rangle_{el}} \nonumber \\
		&+ \mathcal{O}(g_0^3), \label{eq:App_exponential_operator_time_derivative}
	\end{align}
	where we have neglected the commutator $[\dot{\hat{W}}_{\vec{k}, \sigma},  \hat{W}_{\vec{k}, \sigma}(t)]$ and only included terms up to second order in $g_0$. Inserting Eq. (\ref{eq:App_exponential_operator_time_derivative}) into Eq. (\ref{eq:app_time_derivative_full}) yields a term on the form
	\begin{align}
	& \hat{\mathcal{D}}[\beta_{i,i}^{(\vec{k}, \sigma)}(t)] \langle  \dot{\hat{W}}_{\vec{k}, \sigma}(t)  \hat{W}_{\vec{k}, \sigma}(t) \rangle_{el}  ~ e^{-\frac{1}{2} \langle \hat{W}_{\vec{k}, \sigma}^2(t) \rangle_{el}} \nonumber \\
	 & \overset{!}{=}  \langle \langle  \dot{\hat{W}}_{\vec{k}, \sigma}(t)  \hat{W}_{\vec{k}, \sigma}(t) \rangle_{el} 	 \hat{\mathcal{D}}[\beta_{i,i}^{(\vec{k}, \sigma)}(t)] ~ e^{-\frac{1}{2} \langle \hat{W}_{\vec{k}, \sigma}^2(t) \rangle_{el}} , \label{eq:app_commutation}
	\end{align}
	where the equality should be fulfilled by comparing with Eq. (\ref{eq:chi_Markov_W}).
	
	 We note that the equality in Eq. (\ref{eq:app_commutation}) holds if $\hat{\mathcal{D}}[\beta_{i,i}^{(\vec{k}, \sigma)}(t)]$ commutes with $\langle  \dot{\hat{W}}_{\vec{k}, \sigma}(t)  \hat{W}_{\vec{k}, \sigma}(t) \rangle_{el}$. A tedious but straight-forward calculation shows that $\big[\ \hat{\mathcal{D}}[\beta_{i,i}^{(\vec{k}, \sigma)}(t)],\langle  \dot{\hat{W}}_{\vec{k}, \sigma}(t)  \hat{W}_{\vec{k}, \sigma}(t) \rangle_{el} \big] = \mathcal{O}(g_0^3)$ which we disregard due to $g_0 \ll 1$ and the two operators hence commute up to second order in $g_0$. 
	We can thus conclude that the state in Eq. (\ref{eq:chi_Markov_solution}) is a solution to Eq. (\ref{eq:chi_Markov_approx}) up to $g_0^2$. As $e^{-i b(t)}$ is just a phase, it has no physical consequence and is thus ignored.

	\bibliography{this_bib_file}
	
\end{document}